\documentclass[pre]{revtex4-1}

\usepackage{xr-hyper}
\usepackage{amsmath}
\usepackage{amsthm}
\usepackage{amsfonts}
\usepackage{amssymb}
\usepackage{graphicx}
\usepackage[version=4]{mhchem}
\usepackage{mathalfa}
\usepackage{xcolor}

\usepackage[margin=1in]{geometry}
\usepackage[all]{xy}

\newcommand{\one}{($i$) }
\newcommand{\two}{($ii$) }
\newcommand{\three}{($iii$) }
\newcommand{\four}{($iv$) }

\usepackage{mathtools}

\begin{document}

\title{The thermodynamics of quasi-deterministic digital computers.}

\author{
Dominique Chu}

\affiliation{School of Computing, University of Kent, CT2 7NF, Canterbury, UK}
\email{d.f.chu@kent.ac.uk}


\begin{abstract}
A central result of stochastic thermodynamics is that irreversible state transitions of Markovian systems entail a cost in terms of  an infinite entropy production. A corollary of this is that  strictly deterministic computation is not possible. Using a thermodynamically consistent model, we show  that quasi-deterministic computation can be achieved at finite, and indeed modest  cost with accuracies that are indistinguishable from deterministic behaviour for all practical purposes.  Concretely, we  consider the entropy production of  stochastic (Markovian) systems that behave like  AND and a NOT gates.  Combinations of these gates can implement any logical function.   We require that these gates return  the correct result  with  a probability that is very close to 1, and additionally, that  they do so within finite time.  The central component of the model is  a machine that can read  and write binary tapes. We find that the error probability of the computation of these gates  falls with the power of the system size, whereas the cost  only increases linearly with the system size.   
\end{abstract}

\maketitle

\section{Introduction}

There is now a renewed interest in the  statistical mechanics of  information processing \cite{infothermreview}. Research in the area of information thermodynamics focusses on individual processes such as   copying \cite{eule,perscopy}, feedback processes \cite{feedback},   information engines \cite{mandal,mcgrath}, but also computational processes in  chemical systems \cite{government1,myinterfacepaper,wlan}. What has received much less attention is {\em universal computation}, that is processes that can implement arbitrary algorithms, although there has been some efforts  in modelling Turing machines (for example \cite{brandes}), and  references    \cite{wolpertextending,wiese,crutchmandal}  propose general limits on computation but without explicitly relating to specific models of theoretical computer science.
\par
Interest in the physics of computation is not new.   A key result  in the field goes back to the 1980s, stating, somewhat surprisingly, that there is no minimal energy dissipation required during a computation\cite{leffrex,bennetthermo}.  According to this,  computation can be done in principle  at zero energy usage. In practice this zero energy limit is unappealing because it  usually requires  quasi-static processes --- resulting in an infinite computation time --- or it entails an ultra-sensitivity to initial conditions, as for example in the billiard ball computer \cite{billard}.  Complementary to this  is a   more recent result coming out of stochastic thermodynamics stating   that irreversible state transitions in stochastic systems entail  an infinite entropy production. An implication of this is that models of computation that postulate irreversible state transitions,  such as Turing machines or finite state automata,  are physically implausible.  
\par
Real world computing machines must inhabit a regime in-between the infinite dissipation of strictly deterministic machines and the zero-energy limit.  Consistent with this, in biological systems one observes  routinely  trade-offs between the speed, accuracy and energy usage of cellular information processing \cite{myinterfacepaper,wlan,myperformancelimitspaper}. Yet, at the same time,  deterministic computing machines with finite energy dissipation rates do exist. Their existence is not contradicting stochastic thermodynamics because  in reality these machines are not truly deterministic, but they operate at extremely low, practically negligible error rates.  This seems to be sufficient to allow finite, even small, energy dissipation rates of these machines.  
\par
In this contribution we will present a  thermodynamically consistent  model  of   deterministic  computation. By this we mean a computation that \one  returns the ``correct'' result with a probability that is indistinguishable from 1 for all practical purposes, \two does so within finite time, \three is universal. By the latter condition, we mean that the model can be extended so as to implement arbitrary computational functions. \four Finally, we  also assume that the  model is   based on stochastic (Markovian) dynamics. 
\par
 We will focus here on digital, or more specifically, binary computing.   Determinism in analogue computers requires taking the thermodynamic limit, which  leads to poor scaling of cost, accuracy and speed \cite{myperformancelimitspaper}.  More benign scaling can be achieved with digital computation, whereby the state space of the computing machine is partitioned into two equivalence classes. Rather than setting the computer into a specific state, it is only necessary to ensure that the machine is in one of the states of the equivalence class. Thermodynamically, this is much cheaper to achieve.  
\par
 The core element of the  model presented here  are  binary  tapes. Each tape   encodes  a single  bit, corresponding  to the majority of its symbols. The idea here is that the tape represents the record of several attempts to transmit a bit value, whereby each transmission was only successful with some probability $\epsilon$.  A stochastic reading machine  is used to  determine  reliably the bit value represented by the tape.  Variations of such reading machines can mimic NOT and AND gates, and can therefore  be combined to arbitrary logical circuits, thus enabling universal computation. Using this model we will probe  the costs of deterministic computation , both in terms of entropy production and computation time. We  will find that   the scaling of cost and accuracy is benign, conducive to arbitrarily accurate computation at a finite energy expense.  When run in reverse, then the reading machine can be used to write tapes, while drawing work from an external work reservoir. 

\section{Results}

\subsection{The reading machine}
\label{readingmachine}
\begin{figure}
\centering\includegraphics[angle=-0,width=0.3\textwidth]{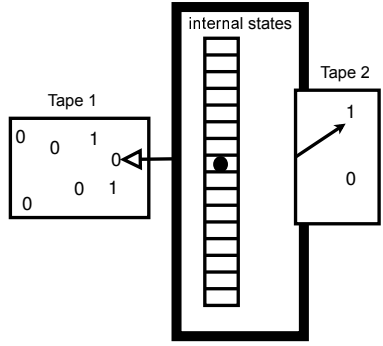}
\caption{The reading machine: The input to the machine is a tape of length $L$ consisting of $m$ symbols "1". The machine has $N+1$ internal states and a display that represents the output tape of length 1.  }
\end{figure}
The central element in the model we propose is the    {\em reading machine} $R$, which is  based on a machine introduced  by  Barato and Seiffert\cite{baratodl1} and thus ultimately on the Mandal-Jarzynski  device \cite{mandal}. The {\em prima facie} function of the machine is to decode a simple repetition error correction code and to set the input to the computation accurately. The machine thus  performs proofreading on  unreliable input.  As will become clear below, $R$ has two further functions: \one  it is   an information processor  for the  computational circuits  and  \two  it also mediates the extraction of free energy   from a ``heat reservoir''  to   power the  computation.   We will first describe how the reading machine works, then determine its accuracy, entropy production and the time its operation takes. Following that, we will show how the reading machine can be used to implement a universal set of logic gates.  
\par
The  machine $R$  interacts with two binary  random access  tapes, $T_1$ and $T_2$, acting as  input and output respectively. By ``random access tape'' we mean that the symbols on the tape are not spatially organised. Each reading event results in a random tape element being accessed.     The input tape $T_1$ is of length $L$ and contains  $m$ copies of the symbol 1  and $(L-m)$ copies of the symbol  0.  The second tape $T_2$ is  of  length 1, i.e.~it is a single bit and  will act  as the output to the machine.  The device also has $N+1$ internal states $s_i^x$; here $x$ is an auxiliary index, indicating the value of $T_2$.  
\par
So as to function as a decoder,  $R$ is to output ``1'' if the majority of bits on $T_1$ is 1, and ``0'' if the majority of bits are symbols of type 0. We do not require that this will work reliably when the  input tape has a slight bias only. However,  the machine must  output the correct bit with probabilities close to 1 for as long as  $\max(m/L,1-m/L) > \theta$  for some fixed  value $1/2<\theta<1$. This can be achived by a  machine that works  according to the following stochastic rules:
\begin{itemize}
\item
At any one time the reading head of the machine accesses (reads)  a symbol of $T_1$.
\item
With rate $k_s$  the reading head  accesses a new symbol of $T_1$.
\item
When  the reading head accesses  a symbol  1 and the internal  state is $s_i^x$ ($i<N$) then  with rate $k_+$  the internal state transitions to $s_{i+1}^x$ and the reading head overwrites the current symbol with a 0.
\item
When  the reading head accesses  a symbol 0 and the internal state is $s_i^x$ ($i>0$)  then with rate $k_-$  the internal state transitions to  $s_{i-1}^x$ and  the reading head overwrites the current symbol with a 1.
\item
When $T_2$ takes the value  0 and the internal state is  $s_N^0$ then with rate $\gamma$ the machine  writes $1$ onto $T_2$ and transitions into internal state  $s_0^1$.
\item
When $T_2$ takes the value 1 and the internal state is  $s_0^1$ then with rate $\gamma$ the machine  writes $0$ onto $T_2$ and transitions into internal state  $s_N^0$.
\end{itemize}
In order to simplify the notation, we will define  a  $\protect\underline{\protect\mathbf 1}$-tape  with respect to  $R$ as a tape of a given length $L$ that, when used as input to $R$,  yields  $T_2=1$ with a steady state probability $\pi(\mathfrak s_{1}) \geq \mathfrak p$. Here we define $\mathfrak s_{1} := s_{j\geq N}$ as the set of states where the output tape $T_2$ is in state 1. Analogously, a  $\protect\underline{\protect\mathbf 0}$-tape  is a tape that, when provided as input to $R$,  outputs 0 with probability   $\pi(\mathfrak s_{0}):= 1-\pi(\mathfrak s_{1})  \geq \mathfrak p$. The parameter  $\mathfrak p$ is a user-defined confidence indicator with $\mathfrak p\approx 1$. There may be  many tapes that are neither $\protect\underline{\protect\mathbf 1}$-tapes nor $\protect\underline{\protect\mathbf 0}$-tapes with respect to a given $R$.  
\par
The behaviour of the  machine $R$ can be modelled as   a random walk  characterised by the rate of interaction with tape-elements $k_s$,   forward rates $k_+ r(m-i\rightarrow m-i-1)$ and backwards rates $k_- r(m-i-1 \rightarrow  m-i)$, where $r(m-i\rightarrow m-i-1)= {m -i\over L} =: \rho_i$ and $r({m-i-1\rightarrow m-i}) = 1 - {m-i-1\over L}:=\bar\rho_{i+1}$. The random walk can be visualised as follows: 
\begin{displaymath}
 \xymatrix{   
m-j+1 & s_{j-1}      & \ldots \ar@{<->}[d]^{k_- \bar\rho_{j-1} }_{k_+ \rho_{j}}  &                         &                         & \\
 m-j  &  s_{j}    & \bullet  \ar@{<->}[r]|{k_s}       & \bullet  \ar@{<->}[d]^{k_- \bar\rho_{j}}_{k_+ \rho_{j+1}}  &                         &  \\   
m-j-1 &  s_{j+1}    &                             &  \bullet  \ar@{<->}[r]|{k_s}  &  \bullet  \ar@{<->}[d] ^{k_- \bar\rho_{j+1}}_{k_+ \rho_{j+2}} &   \\
m-j-2 &   s_{j+2}   &                             &                                &  \bullet  \ar@{<->}[r]|{k_s}                  &   \bullet \ar@{<->}[d] \\  
\vdots      &    \vdots       &                             &                                &                                              &   \vdots  \\  
                                                                         }
\end{displaymath}
The left-most column indicates the number of 1s on the input tape, the second column indicates the internal state and the final column illustrates the transitions. 
For mathematical convenience, but without limiting the generality of our argument, we can assume that the reading head is in a quasi-steady-state with the tape, i.e.~$k_s\gg  k_- \bar\rho_{i},k_+ \rho_{i}$ for all $i$. In this case, the  machine $R$  is simplified to  a 1D random walk on $2N +2$ sites:
%
%
\begin{displaymath}
s_0^0 \longleftrightarrow s_1^0 \longleftrightarrow \cdots \longleftrightarrow {\color{gray} s_N^0 \longleftrightarrow s_0^1} \longleftrightarrow \cdots \longleftrightarrow s_N^1 
\end{displaymath}
The superscript indicates the value of $T_2$.  The transition rates between  states $s^x_i$ and $s^x_{i+1}$ are $k_+ \rho_{i}$ and $k_- \bar\rho_{i+1}$ for the backwards and forward direction respectively, but the rates between $s_N^0$ and $s_0^1$ are $\gamma$  in both  directions. Assuming that $\gamma$ is of the order of the other rates or faster, we can approximate the dynamics of $R$ by cutting out these two sites and connecting $s^0_{N-1}$ directly with $s^1_1$, leading to a random walk on $K:=2N$ sites. As will become clear below, for the parameters of interest, the system spends a vanishing fraction of time on these two sites and the error made by  removing them is minimal. We then end up with the final model, which is a random walk over the states:
\begin{displaymath}
s_0 \longleftrightarrow s_1 \longleftrightarrow \cdots \longleftrightarrow  s_{N-1} \longleftrightarrow s_{N} \longleftrightarrow \cdots \longleftrightarrow s_{K-1} 
\end{displaymath}
Here, the states have been relabelled so that $s_N$ corresponds to $s^1_{1}$ and analogously for other states. The index specifies the number of 1s that have been consumed from $T_1$  in order to reach the specified state.   In the following we will predominantly be interested in the probability $p(\mathfrak s_{1})$   of a particular tape to be recognised as $1$:
\begin{equation}
\label{mainprob}
p(\mathfrak s_{1}) = \sum_{i=N}^{K-1} p(s_i).
\end{equation}
The model makes only sense if $L\geq m\geq K$.

\subsection{Accuracy and resource usage of the reading machine machine}
\label{costs}

In this section we analyse the resource usage of the reading machine. We will find that the ``computation'' time and the entropy production scale linearly with the number of internal state $K$, whereas the error probability scales with the power $-K/2$. 

\subsubsection{Accuracy}

We could now formulate a master equation for the probability  $p_i(t)$  that the system is in state $s_i$ at time $t$; we are however more interested in the corresponding steady-state probability  $\pi_i$. Due to detailed balance the steady-state probabilities obey 
\begin{displaymath}
\pi_i k_+ \rho_{i}  = \pi_{i+1} k_- \bar\rho_{i+1}.
\end{displaymath}
Solving this  for $\pi_i$ yields
\begin{equation}
\label{stateprob}
\pi_i=\eta^{i} 
\underbrace
{\prod_{j=1}^i{\rho_{j}\over \bar\rho_{j+1}}}
_
{{=:u_j}} 
\pi_0, 
\qquad
\eta := {k_+\over k_-}.
\end{equation}
 This leads to an expression of $p_i$ in terms of statistical weights $u_i$
\begin{displaymath}
\pi_i= \eta^i {u_i\over Z}, \qquad\qquad i>0,
\end{displaymath}
where  $Z=\sum_{i=0}^{K-1} \eta^i u_i$ and $\pi_0= 1/Z$. 
\par
\begin{figure}
\centering\includegraphics[angle=-90,width=0.75\textwidth]{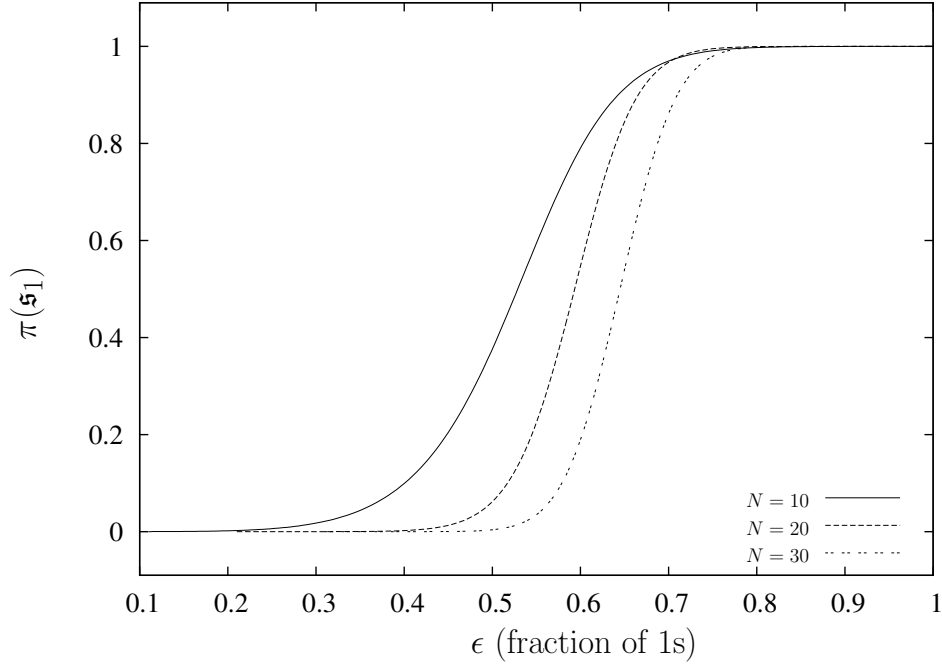}
\caption{The steady-state probability to find the machine $R$  in state $T_2=1$ as a function of $\epsilon$,  the proportion of 1s on the input tape of length $L=100$. The sigmoidal shape of the graph means that for reasonably low/high values of $\epsilon$ the machine outputs 0 and 1 with a probability of almost 1 respectively. }
\label{prob}
\end{figure}
\par
There is no useful analytical expression for this probability, but for  long tapes, when  $L,m\gg K$,  the rates of the random walk are approximately uniform and $m/L$ can be replaced by a fixed fraction $\epsilon$. Remembering that $\mathfrak s_{1} := s_{j\geq N}$ we can now write  the  steady-state probability for $T_2=1$ as
\begin{equation}
\label{probapprox}
\pi(\mathfrak s_{1})\approx
{
\left({\eta \epsilon\over 1 - \epsilon}\right)^{K}
-
 \left({\eta \epsilon\over 1 - \epsilon}\right)^{K\over 2}
\over
\left({\eta \epsilon\over 1 - \epsilon}\right)^{K} -1
}
\end{equation}
This result is exact in the limit $L\rightarrow\infty$.  An important special case for this equation is $\eta=1$ and $K=2$ where the exact and the approximate solutions coincide also for finite $L$ and   the probability to transmit the correct bit becomes $\pi(\mathfrak s_{1})=\epsilon$. This means, that in this case the machine does not improve on the accuracy of the tape, i.e.~it does not perform any proofreading. For $K$ moderately large and $\epsilon>0.5$  eq.~\ref{probapprox} can be  further approximated  to obtain an estimate for the error probability
\begin{equation}
1-\pi(\mathfrak s_{1})\approx\left({\epsilon\over 1-\epsilon}\right)^{-{K\over 2}}.
\label{trude}
\end{equation}
The error falls with the power of $K$. This means  that with a probability that is arbitrarily close to 1 the machine can recognise  tapes correctly even if  $\epsilon$  is only  marginally above  $p=1/2$. When $L\gg K$ then the accuracy of recognition is only limited by  $K$.  For finite $L$ the accuracy $\pi(\mathfrak s_{1})$ increases with $L$ and approaches eq.~\ref{probapprox} asymptotically; the  accuracy also  increases with  $K$, up to an optimal $K$  beyond which the internal mechanism of the machine  deprives the tape of too many 1 symbols and significantly lowers their proportion, which interferes with a proper functioning of the machine.

\subsubsection{Entropy production}

The operation of the reading machine is accompanied by entropy production.  Inserting a tape $T_1$ takes the machine out of equilibrium and initiates a relaxation back to equilibrium. The entropy production ceases on average once equilibrium is reached. Using  the standard {\em ansatz} of stochastic thermodynamics \cite{stochtherm},  the entropy export associated with a transition  from state $i$ to state $k$ is $\Delta s_{\rm env}(i \rightarrow k) = k_{\rm B} \ln \left( \frac{\pi_k}{\pi_i}\right)$. The second component of the entropy is the system or ``Shannon'' entropy which works out as  the difference between the logarithm of the probability of the initial and the final state,  $\Delta s_{\rm sys}(i \rightarrow k) = -k_B \ln\pi_k + k_B \ln p_i(0)$.  The total entropy production is the average  over all initial and final states.  In  the case of reading  a $\protect\underline{\protect\mathbf 1}$-tape  the greatest amount of entropy is produced when   the initial state is  $s_0$, because in this case  the greatest number entropy producing  steps are necessary in order to drive the system to its equilibrium which is a narrow distribution around $s_N$. In this case the entropy production becomes:
\begin{equation}
\label{entropyworst}
\Delta S_{\rm tot}= - k_B   \ln \pi_0.
\end{equation}
In the limit of $L\rightarrow\infty$ an analytic expression for $\pi_0$ can be obtained.
\begin{equation}
\pi_0\approx 
{
(\eta + 1)\epsilon  -1
\over
\left(
\left({\eta \epsilon\over 1 - \epsilon}\right)^{K}
-
1 
\right)
\left( 1-\epsilon\right)
}
\label{approxentr}
\end{equation}
This shows that the entropy production is linear in $K$.

\subsubsection{Computing time}

The second resource consumed by the machine is the  time to reach equilibrium. While the relaxation time is infinite in a strict mathematical sense, a time scale for relaxation can be identified with the mean first passage time (MFPT) to  reach state $s_K$  from some initial state \footnote{We could have equally chosen any site $s_l$ with $l>N$, without altering the conclusions materially.}.  In the worst case, this initial state is $s_0$ in which case  the  MFPT is given by \cite{garten,mylimitedpaper}:
%
\begin{equation}
\label{mfpt}
T_\mathrm{mfpt}=\sum_{i=1}^{K} \sum_{z=0}^{i-1} {1\over k_- \bar\rho_{i-1}} \prod_{l=z+1}^{i-1}{k_+ \rho_{l+1} \over k_- \bar\rho_{l} }
\end{equation}
In general, this formula needs to be evaluated numerically. A compact, albeit approximate analytical expression can be obtained in the case $L,m \gg K$ and $\epsilon = L/m$.
\begin{displaymath}
T_\mathrm{mfpt}\approx
{
\left[\left( (k_- + k_+)\epsilon - k_-   \right)(K+1)   \left(  {\epsilon k_+\over k_- (1-\epsilon)  }  \right)^{K+1} + k_+\epsilon \right]
\left({k_- (1-\epsilon)\over k_+ \epsilon}\right)^{K+1} 
 -k_+ \epsilon
\over
\left((k_- + k_+)\epsilon - k_-\right)^2
}
\end{displaymath}
%
%
For large $K$ and $k_+> k_-$ this  equation can be approximated to  $T_\mathrm{mfpt}\approx (K+1) (Y-k_+)/Y^2 $, where $Y:=(k_-+k_+)\epsilon - k_m$. This shows that the computing time is linear in  $K$. Note that for  finite $L$  the linearity regime is limited to  $K\ll L$. For larger $K$ the time to compute increases exponentially as $K$ grows. Again, the exponential increase is due to the deprivation of the tape for 1 symbols, as $K\rightarrow L$. 
\par
In summary, the reading machine can determine whether a given input $T_1$ is a $\protect\underline{\protect\mathbf 0}$-tape or a $\protect\underline{\protect\mathbf 1}$-tape. By adjusting the parameters of the machine, it is possible to make this decision with arbitrary accuracy at a finite cost and within finite time. 

\subsection{Logic gates}

The reading machine can be used as a basic component to build AND and NOT gates, which in turn can be combined to build arbitrary computational circuits. A   NOT gate  is obtained from the basic reading machine  by swapping the state labels  of $T_2$. This does not affect the properties of the machine, such as the  computing time or the entropy production.   
\par
The AND gate is more involved. It requires two inputs, $A$ and $B$ respectively. We therefore require  an extended reading machine  that accepts two input tapes. Its output is, as in the standard reading machine,   a single element output tape $T_2$.  Each of the inputs $A$ and $B$  can be either a  $\protect\underline{\protect\mathbf 1}$-tape or a $\protect\underline{\protect\mathbf 0}$-tape, each  of length $L$.  Initially $R_\land$ is set up as the combination of two independent, non-interacting, reading machines $R_\land= R_A\otimes R_B$.

The computation of an AND gate proceeds in two separate steps. \one  {\bf Set the input to the gate}. First, the inputs $A$ and $B$ are set by providing each of the independent reading machines $R_A$ and $R_B$ with their respective inputs and letting them reach their equilibrium states.  The  internal states of the combined machine  $R_\land$  can then  be written as   $(s_k^x,r_l)$. Tape $A$  drives state transitions of  type  $(s_k^x, r_l) \rightleftharpoons (s_{k'}^{x'},r_l)$, and $B$ drives interactions of  type $(s_k^x, r_l) \rightleftharpoons (s_{k}^{x},r_{l'})$.  The superscript of the  internal state label $s_k^x$ indicates the bit value  of $T_2$. It  changes from $0$ to $1$ during the transition $(s_{\Theta-1}^0,r_l) \rightleftharpoons (s_\Theta^1,r_l)$, for a fixed threshold $\Theta$ and arbitrary $l$. 
\par
\two {\bf Start the computation proper} after a  time  of order $T_\mathrm{mfpt}$ has passed.  The inputs $A$ and $B$ are then  disconnected and the internal state reservoirs are allowed to interact  by enabling the state transitions    $(s_k^x , r_l) \rightleftharpoons (s_{k+1}^{x'} , r_{l-1})$.
%
The  backwards and forward rates should be equal and independent of $k, x$ and $l$. 
\par
There are choices for the parameters of the reading machines and the threshold $\Theta$ such that the  output  tape $T_2$ behaves  like a quasi-deterministic AND gate.  Define  $M:=k+l$ as  the sum of the indices of $s_k, r_l$ {\em after} the inputs have been set, but before the  internal states are connected. $M$ is  is distributed according to     
\begin{displaymath}
p_\land(M)= \sum_{i = \max(0,M-K+1)}^{\min(M,N-1)} \pi_k \pi_{M-k}. 
\end{displaymath}
If both inputs to the gate are $\protect\underline{\protect\mathbf 1}$-tapes then before the internal states are connected  the state labels of $s_k^x$ and $r_l$ will be the same on average  with  $k \approx l \approx K$. After the computation step this will not change on average. The state will therefore be $(s^1_k,r_l)$, i.e.~the output of the gate is 1 quasi-deterministically  as long as   $\epsilon$ was sufficiently high  on the original tape, and  $\Theta$ is small enough in comparison to $K$. A similar argument applies to the case where both inputs are zero. 
\par
\begin{figure}
\centering\includegraphics[angle=-90,width=0.75\textwidth]{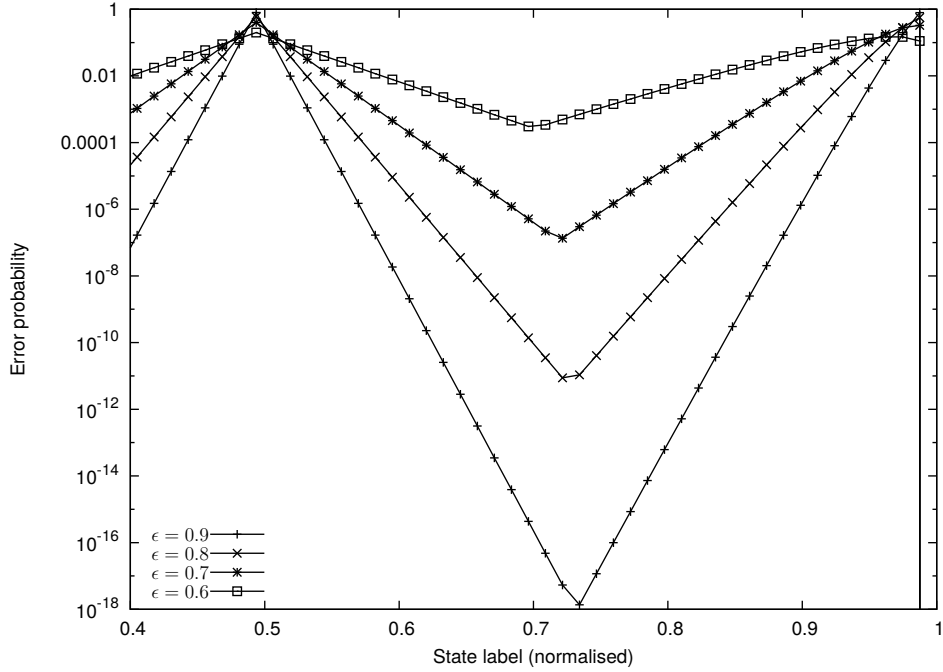}
\caption{Accuracy of the AND gate. The graph shows the $\max(p_\land^{11}(i) ,p_\land^{01}(i))$  for various values of $\epsilon$ and $N=80$. The  threshold $\Theta$ ideally coincides with  a minimum. The probability associated with the minimum also defines the accuracy that can be achieved for the computation.}\label{and}
\end{figure}

The  accuracy of the gate is  limited by the probability to get the  correct output for  mixed input, i.e. a $\protect\underline{\protect\mathbf 1}$-tape and a $\protect\underline{\protect\mathbf 0}$-tape as $A$ and $B$. A correct computation must yield the output 0. Yet,  the average state label of  $s$ after initialisation  will be close to $K$, whereas the average state label of $r$ after initialisation will be close to $0$.  The precise probability distributions for the two cases  are  given by eq.~\ref{stateprob}. After the computation step, both state labels   will be  about $K/2$.  In order for  the output  $T_2$ to be correct,  the threshold  $\Theta$ needs to be chosen such that  the label of state  $s_k$  never  fluctuates to or  beyond $\Theta$ for the  mixed input. Given a set of parameters  $K,L,m$ for the writing machines, there is an  optimal choice for $\Theta$, namely the index  $i>N$ that minimises   $\max(p_\land^{11}(i) ,p_\land^{01}(i))$; see fig.~\ref{and}. Here $p_\land^{11}(i)$ is the probability that the state of the machine is $(s_i^x,r_k)$  given that the input $A$ and $B$ are $\protect\underline{\protect\mathbf 1}$-tapes.
\par
This case of mixed input also leads to an  additional entropy production $\Delta s_\land$, which is  a result of the two internal states being connected and relaxing to a joint equilibrium.   Assume that after setting the input, the machine was in state $(s_k^x,r_l)$ and $M=k + l$.  After  the computation step,  the state  is  $(s_k^x,r_{M-k})$  and the state label $k$ follows a binomial distribution $p(k)={N\choose k}q^k (1-q)^{N-k}$, where  $q:={M\over 2K}$. This  reflects the fact that the internal states have ``equilibrated'' with one another. At the beginning of the computation step, the system is out of equilibrium  with Shannon  entropies $\ln(p(k))$  that are  distributed according to $\pi_k$.  Hence, the change in entropy is   
\begin{equation}
\label{andgate}
\Delta s_\land =  \sum_k \pi_k \ln\pi_k -\sum_k p(k) \ln(p(k)).
\end{equation}
Here, $\pi_i$ is calculated according to eq.~\ref{mainprob}. This entropy production is a direct consequence of the logical irreversibility of the AND gate, and is related to  Landauer's limit. The NOT gate, which is logically reversible, does not have such an extra dissipative component.  The entropy production $\Delta s_\land$ is also the reason why the setting of the input and the computation must be separated processes. If not there would be an ongoing competition between computation and initialisation with ongoing need for energy input.  Note that in the case $A=B$  the  entropy production will be very small, i.e. $\Delta s_\land\approx 0$.  

\subsection{The writing machine}
\label{writingmachine}

A computational cycle is closed by writing the output of the computation to a tape. It is possible to run   the reading machine $R$  run in reverse in order to write a $\protect\underline{\protect\mathbf 1}$-tape or a $\protect\underline{\protect\mathbf 0}$-tape. The following modifications are necessary:  The input tape $T_1$ is  a tape of length 1, the output tape  $T_2$ is  of length $L$ and the machine has  $L+1$ internal states, $s_1, s_2, \ldots, s_{L+1}$. The transition rules of the writing  machine are as follows:
\begin{itemize}
\item
When the internal state is $s_L$ and  $T_1=1$, then with rate $\gamma^w$, the internal state goes into  state $s_0$ while writing $0$ onto $T_1$. The reverse transition happens with the same rate. 
\item
With rate $k_s^\mathrm{w}$ the machine gets in contact with a new symbol from  $T_2$. If the machine is in state $s_i$ then with  probability $({(L-i)/L})$ it will be in contact with  symbol $1$, and with probability $(1-{(L-i)/L})$ it will be  in contact with symbol  0. 
\item
With rate $k^w_-$ the machine writes a $0$ onto the tape and goes into state $s_{i-1}$ (provided $i>0$).
\item
With rate $k_+^w$  the machine writes  a $1$ onto the tape and goes into state $s_{i+1}$ (provided $i<L+1$).  
\end{itemize}
Altogether, the machine, when in state $s_i$ overwrites the current symbol of   $T_2$ with a  $1$ with rate $k_+^w (L-i)/L$ and correspondingly writes a $0$   with rate $k_-^w ( 1-  (L-i)/L  )$. The system can be modelled as a biased random walk of the state label $\iota$. In the long-term limit the average state label has a simple closed form (see SM). 
\begin{equation}
\langle \iota\rangle:=
\sum(i\pi_i^\mathrm{w})= 
L
{
 (\eta^\mathrm{w} + 1)^{L-1} \eta^\mathrm{w})- (\eta^\mathrm{w})^L
\over
1 + (\eta^\mathrm{w} + 1)^L - (\eta^\mathrm{w})^L
} 
\overset{L \gg 1}{\approx}
L{
\eta^\mathrm{w}
\over
\eta^\mathrm{w} + 1
},
\label{avgstate}
\end{equation}
where $\eta^w:=k_+^w/k_-^w$. The accuracy $\epsilon^\mathrm{w}=\langle\iota\rangle/L$ of the writing machine is the probability to find a particular tape element of $T_2$  in the correct state, i.e. a $1$ in the case of a $\protect\underline{\protect\mathbf 1}$-tape or 0 in the case of a $\protect\underline{\protect\mathbf 0}$-tape. 
The  average work required to write a tape with accuracy $\epsilon^\mathrm{w}$ is:  
\begin{equation}
\langle W\rangle:=
k_BTL \epsilon^\mathrm{w} \ln(\eta^\mathrm{w}).
\end{equation}
We can now also relate the error probability of a reading machine to the cost of reconstituting the tape. From eq.~\ref{avgstate}, the   proportion of correct symbols written by the writing machine is $\epsilon^\mathrm{w}\sim \eta^\mathrm{w}/(\eta^\mathrm{w}+1)$; together with eq.~\ref{trude} the  probability that the reading machine fails, then scales  like so
\begin{equation*}
1-\pi(\mathfrak s_{1})\sim\left(\eta^\mathrm{w}\right)^{-{K\over 2}}. 
\tag{\ref{trude}b}
\end{equation*}
%
Finally, the  MFPT to write the full tape can be evaluated along the same lines as  eq.~\ref{mfpt}.  
\begin{displaymath}
T^\mathrm{w}_\mathrm{mfpt}=
\sum_{i=1}^{L}\sum_{x=0}^{i-1}{1\over t^+(i-1)}\prod_{l=x+1}^{i-1}{t^-(l)\over t^+(l)},
\end{displaymath}
where $t^+(l) := k^\mathrm{w}_+ (L-l)/L$ and $t^-(l) := k^\mathrm{w}_- (1-(L-l)/L)$ are shorthand for the forward and backwards rates of the random walk respectively. No useful closed form expression exists for $T^\mathrm{w}_\mathrm{mfpt}$ and it needs to be  evaluated numerically (see SM).  Note that  the computation  time cannot be expressed solely in terms of the ratio $\eta^\mathrm{w}$, but  depends on the absolute scale of the rates. This reflects the fact that the system can be made arbitrarily fast at no additional  cost by scaling the reaction rates. 
\par
The most cost efficient way  to write a tape of a particular type  (i.e. a $\protect\underline{\protect\mathbf 1}$-tape or a $\protect\underline{\protect\mathbf 0}$-tape) with a given accuracy $\epsilon^\mathrm{w}$ is to start from   a relaxed tape of length $L$ with,  on average, $L/2$ symbols of type 1.  To convert this tape into, say, a  $\protect\underline{\protect\mathbf 0}$-tape with (almost) only symbols of 0, only about half the symbols need to be modified.  The average cost of this writing procedure is $-(L/2)k_BT\ln \eta^\mathrm{w}$   with the tape initially in state   $m=L/2$ (see SM). Writing  a $\protect\underline{\protect\mathbf 1}$-tape is entirely analogous, but  requires a special writing machine for $\protect\underline{\protect\mathbf 1}$-tapes. 
\par
Additional costs arise when the bit to be recorded to tape is unknown, which is typically the case at the end of a computational cycle.   One protocol to deal with this case is as follows:
\begin{enumerate}
\item
Prepare a writing machine that outputs $\protect\underline{\protect\mathbf 1}$-tapes with $L$ internal states by resetting its internal state to $s_{L+1}$.
\item
Initialise the input tape $T_1$ of the writing machine with $b$, i.e. the output tape from the preceding computation.
\item
Initialise $T_2$ with a $\protect\underline{\protect\mathbf 0}$-tape.
\item
Wait for a time  of the order  $T^\mathrm{w}_\mathrm{mfpt}$  and then remove  $T_2$. 
\end{enumerate}
If $b=0$ then the machine would not have modified the output tape   $T_2$, no extra cost arises here. Otherwise, the $\protect\underline{\protect\mathbf 0}$-tape  output would have been overwritten to be a $\protect\underline{\protect\mathbf 1}$-tape at a cost proportional to $L$ rather than $L/2$, i.e.  twice the work  to write a $\protect\underline{\protect\mathbf 1}$-tape directly from an initially relaxed tape. In both cases the cost to write the original $\protect\underline{\protect\mathbf 0}$-tape accrues and is $\sim L/2$. 
\par
An additional cost comes from the reset during step 1.  If the machines were not reset, then it  would be initially  in a random state internal state $s_i$, where the state label  $i$ is  uniformly distributed across all possible states  with average  $\langle\iota\rangle = L/2$. This is a source of error, because if the machine is initially in state $s_i$ it would  write $L+1-i$ 1s onto the tape, irrespective of the input. When $b=0$ then this could significantly degrade the quality of the output tape $T_2$. In the case of $b=1$, this would not be harmful though. 
\par
 The reset of the writing machine  comes at the average cost of $\sim L/2$.  Altogether, therefore, the cost of writing a $\protect\underline{\protect\mathbf 1}$-tape  is $\sim 2 L$, which makes the average cost $\sim(3/2)L$.  
\par
This result is not a  fundamental lower limit for the writing of  the output, which can only be reached using quasi-static protocols; also see SM section for alternative ways to copy a tape.

\section{Discussion}
\label{discussion}

In the model presented here, all computational processes complete within a finite time. The error probability  of the computation falls  much more rapidly  to zero than the entropy production increases, which makes it possible to achieve  quasi-deterministic computations at finite cost in finite time.  More specifically, the   entropy production associated with setting the input  diverges linearly with the ability to correct, which is parametrised by   the number of internal states $K$.  This parameter also determines the cost of re-constituting the input tape after the computation. Since in this model the tape serves a dual role as a power source and information storage, the reconstitution cost is the actual cost of the computation. Note that it is normally not necessary to write tapes de-novo at a cost $\sim L/2$ because a computation only overwrites at most $K$ symbols on the input tape.  An additional cost  arises when  executing the AND gate. This  cost  is a consequence of the logical irreversibility of the operation. It too scales linearly with $K$. In contrast to the linear scaling of the cost, the  probability  that the computer returns the wrong results follows  $1 -  \pi(\mathfrak s_1)\sim (\epsilon/(1-\epsilon))^{-K/2}$ (see eq.~\ref{trude}). 
\par
The benign scaling  of this machine is re-assuring vis-\`a-vis  the existence of real world deterministic computing  machines, which are  in reality  only quasi-deterministic, i.e.~ stochastic with a  very low error probability. Indeed,  determinism in electronic circuits is achieved by using principles that are  formally not too dissimilar from the model presented here.   Bit values are represented as voltage spikes. If the  amplitude of a spike  exceeds a certain voltage threshold, then it is interpreted as a 1.  The probability of an error can be reduced arbitrarily by choosing the correct threshold value in relation to the average voltage peak and typical fluctuations.    
\par
All this begs the question why biological systems do not, at least not universally,  use a similar route to deterministic computation. Unlike electronic machine, {\em in vivo} computation is inherently stochastic and  subject to performance trade-offs.  Part of the explanation may be that cellular computing is analogue, rather than digital, and not admitting such a benign scaling. Another  reason could be that the infra-structure required to perform digital computation cheaply, is itself not cheap to maintain. Here, we have not  included the maintenance  cost  of the reading machine whereas a  biological cell has to  produce and maintain or ``compute'' the reading machine itself. This may not be worthwhile doing. 
\par
As a final remark, we note that  the model used here  is but an application of the idea of  non-confusable subset  coding from information theory \cite{mckayinfo}.  One may wonder   whether  more advanced block coding schemes could be used to get an  even   better performance. We conjecture that  the computational cost of decoding puts a limit to the  use of error correction codes.  The computation necessary during the decoding step would itself require energy and likely render the energy-accuracy balance unfavourable. 

\begin{acknowledgments}

The author thanks Thomas Ouldridge for valuable comments and discussions on early drafts of this manuscript. 
\end{acknowledgments}


\end{document}